# Pseudo spin torque induced by strain field of Dirac fermions in graphene

**Bumned Soodchomshom**[a,*]

[a] Department of Physics, Faculty of Science, Kasetsart University Bangkok 10900, Thailand

**Abstract**

Physical property of pseudo spin of electrons in graphene is investigated. In contrast to recent description [Phys. Rev. Lett. 106 (2011)116803], we show that pseudo spin in graphene is not completely a real angular momentum. The pseudo spin only in the direction perpendicular to graphene sheet is real angular momentum, while the pseudo spin parallel to graphene plane is still not real angular momentum. Interestingly, it is also shown that the Newtonian-like force and pseudo spin torque of massive Dirac electrons in graphene under strain field mimic gravitomagnetic force and gravitomagnetic spin torque, respectively. This is due to the equivalence of pseudo spin and velocity operators of 2+1 dimensional massive electrons in graphene, different from that in real 3+1 dimensional Dirac fields. This work reveals new physical property of graphene as a pseudo gravitomagnetic material.

## 1. Introduction

Two-dimensional Dirac fermions in graphene [1, 2] carrying pseudo spin generated by its two-sublattices has lead to a way that would connect condensed matter and high energy physics. Such pseudo spinnor field is associated with hopping motions of electron in honeycomb lattice of graphene atomic structure. Analogous to that in high energy system, the emergence of pseudo Dirac-like field gives rise to Klein tunneling which is due to conservation of pseudo spin, experimentally observed in graphene p-n junction [3]. Zitterbewegung, trembling motion of relativistic fermions, may be studied via electrons in graphene system [4-5]. Discovery of pseudo spinnor field in graphene may also lead to the understanding of physics behind the theory of relativity, since electrons in graphene possess pseudo relativistic fermions, despite low energy. In graphene, the Fermi energy plays the role of the speed of light. Holes and sublattices are considered as pseudo positrons and pseudo spin, respectively. The artificial spin, pseudo spin, created by the presence of the two-sublattices in graphene is one of the exotic natures that may be used to understanding the relationship between spin and orbital angular momentum. Some previous works [6, 7] asserted that lattice-induced pseudo spin in graphene may not be associated with angular momentum, while a recent work [8] has given a different description. Mechlenburg and Regan [8] has shown that projection of total angular momentum $\vec{J} = \vec{L} + \vec{S}$ onto the normal direction ($J_\perp$) of graphene sheet is a conserved quantity, ie.,

$$[\hat{H}, J_\perp] = \vec{0}, \qquad (1)$$

where $\hat{H}$ is Hamiltonian of electron in graphene. $\vec{L} = \vec{r} \times \vec{p}$ is the orbital angular momentum of electron in graphene with $\vec{r}$ and $\vec{p}$ being position and momentum vectors, respectively. $\vec{S}$ is pseudo spin operator of electron in graphene. This result may lead their conclusion to that pseudo spin in graphene is a real angular momentum. In this letter, we will show that the proof of Ref.[8] is not enough for the conclusion that pseudo spin in all directions is a real angular momentum because the in-plane total angular momentum ($J_\parallel$) in Ref.[8] is not a conserved quantity, ie.,

$$[\hat{H}, J_\parallel] \neq \vec{0}. \qquad (2)$$

The proof in Ref.[8] should be sufficient only for the pseudo spin perpendicular to graphene sheet to conclude that it is a real angular momentum, not including the in-plane pseudo spin. The physical nature of the in-plane pseudo spin is still unclear. In this work, we show that the in-plane pseudo spin parallel to graphene sheet is not real angular momentum, in contrast to Ref.[8]. We also show that the Newtonian-like force and pseudo spin torque of massive Dirac electrons in graphene under strain field are analogous to gravitomagnetic force and gravitomagnetic spin torque, respectively. The pseudo spin generating spin gravitomagnetic-like moment, leading to graphene as a pseudo gravitomagnetic material.

**2. To recall the orbital angular momentum and spin of the real 3+1 dimensional Dirac fermions**

The classical torque $\vec{\tau}_C$ defined as $\vec{\tau}_C = \frac{d\vec{L}}{dt} = \frac{d(\vec{r} \times \vec{p})}{dt} = \frac{d\vec{r}}{dt} \times \vec{p} + \vec{r} \times \frac{d\vec{p}}{dt}$. In the quantum mechanics, via the Heisenberg's picture, the orbital torque operator $\vec{\tau}_Q$ may be rewritten as of the form

$$\vec{\tau}_Q = \frac{i}{\hbar}\left[\hat{H}, \vec{r}\right] \times \vec{p} + \vec{r} \times \frac{i}{\hbar}\left[\hat{H}, \vec{p}\right] = \vec{v} \times \vec{p} + \vec{r} \times \vec{F},$$

(3)

where velocity operator and force operator are given by $\frac{d\vec{r}}{dt} = \frac{i}{\hbar}\left[\hat{H}, \vec{r}\right] = \vec{v}$ and $\frac{d\vec{p}}{dt} = \frac{i}{\hbar}\left[\hat{H}, \vec{p}\right] = \vec{F}$, respectively. In nonrelativistic free particle with mass "m", Schrödinger Hamiltonian $H_{sh} = \frac{\vec{p}^2}{2m}$ leads to the result that

$$\vec{\tau}_{Sh} = \frac{i}{\hbar}\left[\hat{H}_{Sh}, \vec{r}\right] \times \vec{p} + \vec{r} \times \frac{i}{\hbar}\left[\hat{H}_{Sh}, \vec{p}\right] = \vec{0}. \quad (4)$$

The angular momentum is a conserved quantity, because velocity always parallel to momentum $\vec{v}_{Sh} \times \vec{p} = 0$ in the classical model. The spin angular momentum can not be mathematically observed in case of non relativistic model. The total angular momentum of the free particle thus belongs only to the orbital angular momentum $\vec{L}$.

Using eq.(3) to study the orbital torque operator $\vec{\tau}_D$ in the free field of the real 3+1 Dirac fermions with the Dirac Hamiltonian $\hat{H}_D = c\vec{\alpha}\cdot\vec{p} + \beta mc^2$ where $\vec{\alpha} = \begin{pmatrix} 0 & \vec{\sigma} \\ \vec{\sigma} & 0 \end{pmatrix}$ and $\beta = \begin{pmatrix} 1 & 0 \\ 0 & -1 \end{pmatrix}$ with $\vec{\sigma} = \sigma_x \hat{i} + \sigma_y \hat{j} + \sigma_z \hat{k}$ being the vector of the Pauli spin matrices. The speed of light is "c". The unit vectors along the x, y and z axis are $\hat{i}, \hat{j}$ and $\hat{k}$, respectively. The orbital torque operator for the free 3+1 dimensional Dirac field may be given as

$$\vec{\tau}_D \approx \frac{d\vec{L}}{dt} = \frac{i}{\hbar}\left[\hat{H}_D, \vec{r}\right] \times \vec{p} + \vec{r} \times \frac{i}{\hbar}\left[\hat{H}_D, \vec{p}\right] = \vec{v}_D \times \vec{p}, \quad (5)$$

where the Dirac velocity operator is $\vec{v}_D = \frac{i}{\hbar}\left[\hat{H}_D, \vec{r}\right] = c\vec{\alpha}$. The force operator vanishes for the free particle $\frac{i}{\hbar}\left[\hat{H}_D, \vec{p}\right] = 0$, although there is non-vanishing acceleration $\frac{d\vec{v}_D}{dt} = \frac{i}{\hbar}\left[\hat{H}_D, \vec{v}_D\right] \neq \vec{0}$ leading to the so-called "Zitterbewegung". This is to say the quantum mechanics may allow $\vec{v} \times \vec{p} \neq \vec{0}$ to cause Zitterbewegung for some systems in the condensed matter [9], giving rise to orbital torque generated without the external force $\vec{F}$. As seen in eq.(5), the angular momentum is not a conserved quantity. The total angular momentum $\vec{J}_D = \vec{L} + \vec{S}_D$ is required and should be a conserved quantity ie., $\frac{d\vec{J}_D}{dt} = \frac{i}{\hbar}\left[\hat{H}_D, \vec{L} + \vec{S}_D\right] = \vec{0}$. Because the condition $\frac{d\vec{S}_D}{dt} = -\frac{d\vec{L}}{dt} \neq \vec{0}$ is required [10], the spin angular momentum operator $\vec{S}_D$ must be defined to obey the relation

$$\frac{i}{\hbar}\left[\hat{H}_D, \vec{S}_D\right] = -\frac{i}{\hbar}\left[\hat{H}_D, \vec{L}\right] = -\frac{i}{\hbar}\left[\hat{H}_D, \vec{r}\right] \times \vec{p} = -\vec{v}_D \times \vec{p}.$$

(6)

From eq.(6), real spin operator of the real 3+1 dimensional Dirac fermions is given as $\vec{S}_D = \frac{\hbar}{2}\begin{pmatrix} \vec{\sigma} & 0 \\ 0 & \vec{\sigma} \end{pmatrix}$. The formalism in eq.(6) should be the standard form of the relationship among **spin**, **orbital angular momentum**, **position** and **velocity operators** for free fields of the Dirac-like fermions. We would like to conclude that the pseudo spin of the pseudo 2+1 dimensional Dirac fermions in graphene must be defined using this form because it is a Dirac-like fermions.

As we have mentioned previously, Ref.[8] did not obey the relation of $\frac{d\vec{J}}{dt} = \frac{i}{\hbar}\left[\hat{H}, \vec{L} + \vec{S}\right] = \vec{0}$. Hence, the proof may be incomplete, except for spin in the z direction or $S_\perp$ (see the direction defined in Fig.1). In the next section we will consider the physical nature of pseudo spin in graphene based on the relation given in eq.(6) and show the different result from Ref.[8].

**3. Property of the pseudo spin of 2+1 dimensional Dirac fermions in graphene**

The tight-binding based electronic fields in graphene are described as pseudo 2+1 dimensional Dirac fermions [1]. Using the coordinates xyz defined with respect to the real space of the graphene atomic structure (see Fig.1), the valley-dependent Hamiltonian acting on the pseudo spinnor field $\psi \sim \begin{pmatrix} \psi_A \\ \psi_B \end{pmatrix}$ generated by sublattices A and B for the free electrons taking the form [11]

$$\hat{H}_{G,\kappa} = v_F\left(\kappa\sigma_x p_x + \sigma_y p_y\right) + \kappa\Delta\sigma_z \qquad (7)$$

where, $v_F \approx 10^6$ m/s is the Fermi velocity and the valley $K(K')$ is denoted by $\kappa = 1(-1)$. The energy gap between valence and conduction band is $2\Delta$ and may be created by growing graphene on hexagonal boron nitride [12, 13]. The energy gap opening behaves like mass of the Dirac fermions called "pseudo Dirac mass" $m_{pseudo} \approx \Delta/v_F^2$. Recently, we have shown that pseudo Dirac mass can be either positive or negative value [13]. In the graphene system, the pseudo spin operator (or spin-like operator) may be defined as a form of operator that leads to Eigenvalue of $\hbar/2$ (fermions with spin 1/2)

$$\vec{S}_{G,\kappa} = \frac{\hbar}{2}\left(\kappa\sigma_x\hat{i} + \sigma_y\hat{j} + \kappa\sigma_z\hat{k}\right) \qquad (8)$$

When pseudo spin operator has been defined as given in eq.(8), the time evolution of pseudo spin operator is therefore obtained as

$$\frac{d\vec{S}_{G,\kappa}}{dt} \approx \frac{i}{\hbar}\left[\hat{H}_{G,\kappa}, \vec{S}_{G,\kappa}\right] = -\left(\kappa v_F \sigma_x \hat{i} + v_F \sigma_y \hat{j} + \kappa v_F \sigma_z \hat{k}\right) \times \left(p_x \hat{i} + p_y \hat{j} + m_{pseudo} v_F \hat{k}\right)$$

$$= -\vec{v}_{G,\kappa} \times \vec{p}_G.$$

(9)

To compare eq.(9) with eq.(6), we may define **velocity-like operators** $\vec{v}_{G,\kappa}$, **momentum-like operator** $\vec{p}_G$, and **orbital angular momentum-like operator** $\vec{L}_{G,\kappa}$ as of the forms

$$\vec{v}_{G,\kappa} = \frac{i}{\hbar}\left[\hat{H}_{G,\kappa}, \vec{r}_G\right] = \kappa v_F \sigma_x \hat{i} + v_F \sigma_y \hat{j} + \kappa v_F \sigma_z \hat{k},$$

$$\vec{p}_G = p_x \hat{i} + p_y \hat{j} + m_{pseudo} v_F \hat{k}.$$

and $\quad\vec{L}_{G,\kappa} = \vec{r}_G \times \vec{p}_G$, respectively. (10)

This leads to

$$\frac{d\vec{L}_G}{dt} = \frac{i}{\hbar}\left[\hat{H}_{G,\kappa}, \vec{L}_G\right] = \vec{v}_G \times \vec{p}_G, \text{ and } \frac{d\vec{J}_{G,\kappa}}{dt} = \frac{i}{\hbar}\left[\hat{H}_{G,\kappa}, \vec{J}_{G,\kappa}\right] = \vec{0},$$

(11)

where, $\vec{J}_{G,\kappa} = \vec{L}_{G,\kappa} + \vec{S}_{G,\kappa}$ is a **total angular momentum-like operator** which plays the role of a constant total angular momentum of the pseudo 2+1 dimensional Dirac fermions.

In eq.(10), the space of pseudo Dirac fermions in graphene may be described by a 3-dimensional-like system where $\vec{r}_G = \vec{r} + \vec{r}_{pseudo}$ with $\vec{r} = x\hat{i} + y\hat{j}$. From eq.(10), the operator of the positions should be satisfied the conditions of

$$\vec{v} = \frac{i}{\hbar}\left[\hat{H}_{G,\kappa}, \vec{r}\right] = v_F\left(\kappa\sigma_x \hat{i} + \sigma_x \hat{j}\right) \text{ and } \vec{v}_{pseudo} = \frac{i}{\hbar}\left[\hat{H}_{G,\kappa}, \vec{r}_{pseudo}\right] = v_F\left(\kappa\sigma_z \hat{k}\right).$$

(12)

Therefore the operator $\vec{r}_{pseudo}$ may be defined by

$$\vec{r}_{pseudo} = z_{pseudo}\hat{k} = -i\frac{\hbar}{v_F}\frac{\partial}{\partial m_{pseudo}}\hat{k} \ . \qquad (13)$$

The result we have obtained in our work now shows that the pseudo 2+1 dimensional Dirac fermions act like particle moving in 3-dimensional space with a **position-like vector**, $\vec{r}_G = x\hat{i} + y\hat{j} + z_{pseudo}\hat{k}$.

Now considering eq.(11), it is seen that pseudo spin $\vec{S}_{G,\kappa}$ is associated with the (operator-like) orbital angular momentum $\vec{L}_{G,\kappa} = \vec{r}_G \times \vec{p}_G$. Here we would say that pseudo spin operator "$\vec{S}_{G,\kappa}$" behaves like a real angular momentum only in the z-direction, because it is associated with real orbital angular momentum $L_{G,z} = xp_y - yp_x$, in contrast to the xy-plane. $\vec{S}_{G,xy}$ should be not associated with any real angular momentum (see table 1), because it is related to pseudo orbital angular momentum

$$L_{G,x} = yp_{pseudo} - z_{pseudo}p_y \text{ and } L_{G,y} = z_{pseudo}p_x - xp_{pseudo} \ .$$

This is due to that effective electronic field in graphene is actually confined only in the xy-plane. Hence, our work shows a different conclusion from previous work [8], on the description of physical property of pseudo spin or lattice spin in graphene.

**3. Pseudo spin torque due to gravitomagnetic-like field generated by strain**

In graphene system, pseudo spin operator $\vec{S}_{G,\kappa}$ is found to resemble velocity-like operator $\vec{v}_{G,\kappa} \approx \frac{2v_F}{\hbar}\vec{S}_{G,\kappa}$ (see table 2 and eq.(12)). Pseudo spin is parallel to the direction of velocity. This property is one of the interesting behaviors of pseudo 2+1 dimensional Dirac fermions in graphene, unlike that in 3+1 dimensional Dirac fermions. In this section, we focus on the pseudo spin torque and acceleration force that may be related to Zitterbewegung of electrons in graphene where the strain field is applied. Under the influence of strain field [11], the graphene Hamiltonian may be given as

$$\hat{H}'_{G,\kappa} = \kappa v_F \sigma_x \left( p_x + \kappa A_x \right) + v_F \sigma_y \left( p_y + \kappa A_y \right) + \kappa m_{pseudo} v_F^2 \sigma_z ,$$

(14)

where $\vec{A}_{strain} \approx A_x \hat{i} + A_y \hat{j}$ is defined proportional to a pseudo vector potential related to a pseudo magnetic field $\vec{B}_{mag} \propto \vec{\nabla} \times \vec{A}_{strain}$ which is due to perturbed hoping energies induced by strain [4, 14]. Under strain-induced vector potential, pseudo spin torque $\vec{\tau}_{ps} \approx \dfrac{d\vec{S}_{G,\kappa}}{dt} = \dfrac{i}{\hbar}\left[ \hat{H}'_{G,\kappa}, \vec{S}_{G,\kappa} \right]$ could be obtained as

$$\vec{\tau}_{ps} = -\frac{2v_F}{\hbar}\vec{S}_{G,\kappa} \times \vec{p} - \kappa \frac{2v_F}{\hbar}\vec{S}_{G,\kappa} \times \left( \vec{A}_{strain} + m_{pseudo} v_F \hat{k} \right).$$

(15)

To consider the case of rest electrons $\vec{p} = p_x \hat{i} + p_y \hat{j} \to \vec{0}$, pseudo spin torque is purely induced by strain and mass and we may obtain the formalism in eq. (15) as of the form

$$\vec{\tau}_{ps} = \vec{S}_{G,\kappa} \times \left( \frac{-2\kappa v_F}{\hbar}\left( \vec{A}_{strain} + m_{pseudo} v_F \hat{k} \right) \right).$$

(16)

Previously [1, 4, 11, 14], it has been understood that strain field in graphene may create pseudo magnetic filed. The spin torque found in eq.(16) is to show that strain vector field $\vec{A}_{strain}$ acts like a component of a magnetic-like field rather than a vector potential. In this section, we will show that strain field could also create a gravitomagnetic-like field.

As we have known, the inertia force is defined as a derivative of momentum respected to time $\vec{F} \approx \dfrac{d\vec{p}}{dt}$. Considering the case of graphene under the influence of a "constant strain field", the inertia-like force is zero $\vec{F}_{G,\kappa} \approx \dfrac{i}{\hbar}\left[ \hat{H}'_{G,\kappa}, \vec{p}_G \right] = \vec{0}$, while its acceleration does not vanishes $\vec{a}_{G,\kappa} \approx \dfrac{i}{\hbar}\left[ \hat{H}'_{G,\kappa}, \vec{v}_{G,\kappa} \right] \neq \vec{0}$. It is not unclear that this acceleration appears without being induced by related external force. We may define the force inducing this acceleration by using **Newtonian-like second law,** given by

$$\vec{f}_{G,\kappa} \approx m_{pseudo}\vec{a}_{G,\kappa} = \frac{i}{\hbar}\left[\hat{H}'_{G,\kappa}, m_{pseudo}\vec{v}_{G,\kappa}\right]$$

$$= -m_{pseudo}\vec{v}_{G,\kappa} \times \left(\frac{2v_F\vec{p}}{\hbar}\right) - \kappa m_{pseudo}\vec{v}_{G,\kappa} \times \left(\frac{2v_F}{\hbar}(\vec{A}_{strain} + m_{pseudo}v_F\hat{k})\right).$$

(17)

To take $\vec{p} = p_x\hat{i} + p_y\hat{j} \to \vec{0}$ into eq. (17), we may get the force-induced acceleration taking the form

$$\vec{f}_{G,\kappa} = m_{pseudo}\vec{v}_{G,\kappa} \times \left(\frac{-2\kappa v_F}{\hbar}(\vec{A}_{strain} + m_{pseudo}v_F\hat{k})\right).$$

(18)

Recalling electromagnetic formulism, we will get spin magnetic torque and magnetic force-induced cyclotron motion as of the forms $\vec{\tau}_{mag} = \vec{\mu}_{mag} \times \vec{B}_{mag}$ and $\vec{f}_{mag} = q\vec{v} \times \vec{B}_{mag}$, respectively where $\vec{\mu}_{mag} \approx g_M\frac{e}{2m}\vec{S}$ is electron spin magnetic moment with spin gyro magnetic ratio $g_M \approx 2$, q is electronic charge, and $\vec{B}_{mag}$ is a magnetic field strength.

To consider formulae in eqs.(16) and (18), they can be rewritten similar to the formalisms of "**gravitomagnetic spin torque**" and "**gravitomagnetic force**" [15, 16], given as

$$\vec{\tau}_{ps} = \vec{\mu}_{G,\kappa} \times \vec{B}_{G,\kappa}, \text{ and } \vec{f}_{G,\kappa} = m_{pseudo}\vec{v}_{G,\kappa} \times \vec{B}_{G,\kappa},$$

(19)

respectively, where $\vec{\mu}_{G,\kappa} \approx \vec{S}_{G,\kappa}$ and $\vec{B}_{G,\kappa} \approx \frac{-2\kappa v_F}{\hbar}(\vec{A}_{strain} + m_{pseudo}v_F\hat{k})$ could be considered as "**spin gravitomagnetic-like moment**" and "**gravitomagnetic-like field strength**", respectively. It may be easy to approximate the orbital angular momentum inducing gravitomagnetic moment defined as $\mu_{Gr} = I_m A$ by classical trajectory in the xy plane, where mass current is "$I_m$" and the area of the loop current is "$A$". Thus, we take $mvr = L_z$ with $v \approx \frac{2\pi r}{t}$ being velocity. Hence, gravitomagnetic moment is found to be proportional to the angular momentum ie., $\mu_{Gr} = I_m A \approx \frac{L_z}{2} \approx$ angular momentum/2 [16], where the mass current $I_m \approx \frac{m}{t}$ and the area $A \approx \pi r^2$. Therefore, spin gravitomagnetic moment could be written as $\vec{\mu}_{pseudo,\kappa} = \vec{S}_{G,\kappa} \approx g_{Gr} \frac{1}{2} \vec{S}_{G,\kappa}$ leading to gyro gravitomagnetic-like ratio approximated of $g_{Gr} \approx 2$ similar to real spin gyro magnetic ratio of bare electron.

In a limiting case for perfect graphene, pseudo Dirac electrons are massless $\Delta = m_{pseudo} v_F^2 \to 0$ due to its zero-band gap. Hence, the gravitomagnetic-like field is generated only by the strain filed $\vec{B}_{G,\kappa} \approx \frac{-2\kappa v_F}{\hbar} \vec{A}_{strain}$. Although the Newtonian-like force vanishes $\vec{f}_{G,\kappa} \approx \vec{0}$ in perfect graphene, the acceleration always remains finite $\vec{f}_{G,\kappa} / m_{pseudo} = \vec{a}_{G,\kappa} \approx \vec{v}_{G,\kappa} \times \vec{B}_{G,\kappa} \neq 0$. The result in eqs.(19) may be still done in case of perfect graphene. Force-induced cyclotron-like trajectory is replaced by the acceleration formula $\vec{f}_{G,\kappa} / m_{pseudo} = \vec{a}_{G,\kappa} \approx \vec{v}_{G,\kappa} \times \vec{B}_{G,\kappa} \neq 0$. The perfect graphene with zero-band gap is just the limiting case of our result.

## 4. Summary and conclusion

Physical properties of pseudo spin of pseudo 2+1 Dirac fermions in graphene has been studied. It has been found that in contrast to recent description [8], the pseudo spin of electrons in graphene generated by its sublattices is not completely a real angular momentum. The pseudo spin parallel to graphene plane could not be described as a real angular momentum, except for normal component of the pseudo spin. Interestingly, the gravitomagnetic-like field induced by strain for massive Dirac fermions in graphene system was predicted, due to the equivalence of pseudo spin and velocity operators of 2+1 dimensional massive electrons in graphene, different from real 3+1 dimensional Dirac fields. The pseudo spin was found to generate spin gravitomagnetic-like moment. This work has revealed new physical property of strained graphene as a pseudo gravitomagnetic material. A new result may leads to turn graphene into pseudo gravity physics.

**Acknowledgements**

This work was supported by Kasetsart University Research and Development Institute (KURDI) and Thailand Research Fund (TRF) under Grant. No. MRG5580237.

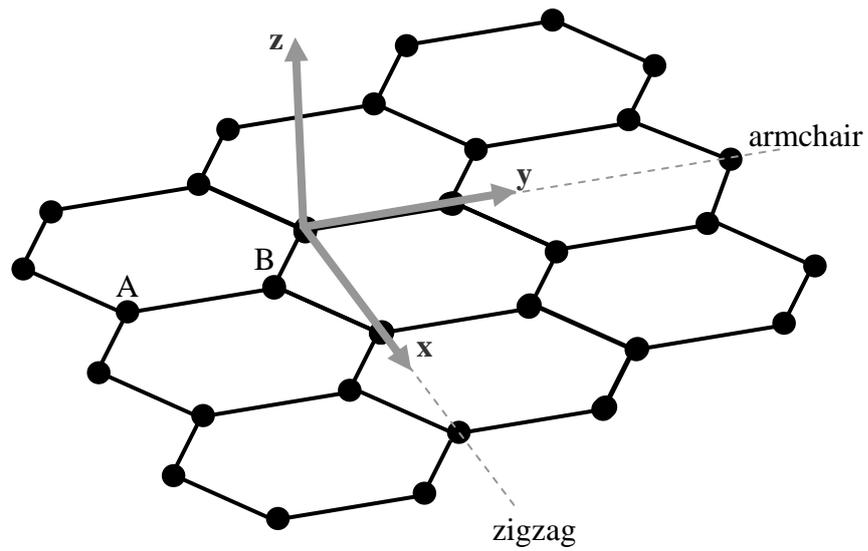

**Figure 1** illustration of graphene atomic structure and directions of x-, y- and z- directions. The zigzag and armchair directions are parallel to x- and y- directions respectively. The z-direction is perpendicular ($\perp$) to graphene plane. Real trajectory of electrons in graphene is confined in the xy-plane ($\parallel$).

| $\vec{S}$ | $\vec{L}_G = \vec{r}_G \times \vec{p}_G$ | $\vec{J}_{G,\kappa} = \vec{L}_{G,\kappa} + \vec{S}_{G,\kappa}$ |
|---|---|---|
| $S_{G,x} = \kappa \dfrac{\hbar}{2}\sigma_x$ | $L_{G,x} = yp_{pseudo} - z_{pseudo}p_y$ | **Not real** angular momentum |
| $S_{G,y} = \dfrac{\hbar}{2}\sigma_y$ | $L_{G,y} = z_{pseudo}p_x - xp_{pseudo}$ | **Not real** angular momentum |
| $S_{G,z} = \kappa \dfrac{\hbar}{2}\sigma_z$ | $L_{G,z} = xp_y - yp_x$ | **Real** angular momentum |

**Table 1** shows properties of pseudo spin $\vec{S}_{G,\kappa}$, orbital angular momentum-like operator $\vec{L}_{G,\kappa}$ and total angular momentum like operator $\vec{J}_{G,\kappa}$. The momentum-like operator $p_{pseudo} \approx m_{pseudo}v_F$ is defined related to pseudo motion in the z direction.

| $\vec{r}_G$ | $\vec{v}_{G,\kappa} = \dfrac{2v_F}{\hbar}\vec{S}_{G,\kappa}$ | $\vec{p}_G$ | property |
|---|---|---|---|
| x | $\dfrac{i}{\hbar}\left[\hat{H}_{G,\kappa}, x\right] = v_F \kappa \sigma_x$ | $p_x \approx -i\hbar\dfrac{\partial}{\partial x}$ | **Real** |
| y | $\dfrac{i}{\hbar}\left[\hat{H}_{G,\kappa}, y\right] = v_F \sigma_y$ | $p_y \approx -i\hbar\dfrac{\partial}{\partial y}$ | **Real** |
| $z_{pseudo} \approx -i\dfrac{\hbar}{v_F}\dfrac{\partial}{\partial m_{pseudo}}$ | $\dfrac{i}{\hbar}\left[\hat{H}_{G,\kappa}, z_{pseudo}\right] = v_F \kappa \sigma_z$ | $m_{pseudo} v_F$ | **Not real** |

**Table 2** shows properties of position-like operator $\vec{r}_G$, velocity-like operator $\vec{v}_{G,\kappa}$ and momentum-like operator $\vec{p}_G$.